# Electric Democracy: Proof of Work to secure Elections

By *Vitaly Zuevsky*[1]

**Abstract**. Electronic voting consistently fails to supplant conventional paper ballot due to a plethora of security shortcomings[2,3]. Not only are traditional voting methods mediocre in terms of convenience and interface, they also encompass principal-agent problem, where the state may have vested interest in the outcome of the ballot. Electronic voting protocol using cryptography to deliver in zero trust environments is long overdue. Here I propose using Proof of Work algorithm[5] at user devices in combination with other well-known security primitives to build a zero-trust voting system. The state would only issue single-use voting authorizations to its citizens, while zero trust design would allow conducting elections by an open-source platform with enhanced observability. It is also hypothesized that the heighten availability of plebiscite by means of the proposed design may ultimately change the way our society participates in the policy making.

Democracy is a great achievement of our evolution, but its implementation is prone to fraud. There is an inherent conflict of interest, where a state authority manages an election, because the ruling cohort has the real capabilities to tamper with the outcome. We read reports about fraudulent actors past every election in the first world's nations, leaving the scale of the problem to a wild guess in less developed – more corrupted counties.

The problem has little prospect of solution in principle as long as the voting process remains on paper. There appears to be a growing number of electronic alternatives[2], but they lack inherent guarantees against tampering. There were also a number of blockchain based attempts at voting, although they were all focused on blockchain per se as opposed to a secure design for elections[3].

There are two main challenges in electronic voting systems today. First of all, as voter's eligibility must be guaranteed and their vote can be casted only once, whoever owns the voting system sees voters' votes as they come in, breaking the principle of secrecy. Lack of secrecy or even mere possibility thereof can influence votes being casted, rendering elections susceptible to propaganda and oppressive ideology.

From cryptology standpoint eligibility guarantee can be formalized as zero knowledge proof for set membership, where the set is a list of all eligible voters. Although such corpus of work exists[4], it is not sufficiently scalable at present. Moreover, zero knowledge proof of voter eligibility would make it difficult to guarantee that the vote was casted only once. The only practical solution today is removing state agency altogether and, instead, using an external state-independent entity to conduct elections and to publish two separated lists: the voted voters and the votes. Even though the voter can be potentially identified from the first list, there will be no linking to their vote in the second list. The entity conducting the election severs that link.

Here we are coming to the second challenge. How could it be guaranteed that the lists in question are not tampered with during all that processing and publication, especially in the circumstances where neither a state nor its citizens trust each other, let alone external entities they are to entrust elections? It turns out Proof of Work technique[5] powering blockchain technology but not blockchain itself can help.

As more computationally capable devices – such as smartphones, tablets, and laptops – penetrate our population, they can collectively create a sufficient body of computational work to serve as a deterrent against tampering. That would at least hold true for tampering with a significant share of votes during a limited period of time an election is taking place for.

Publishing of the lists is supposed to happen when the election is over. It is also possible to publish in chunks to make it more of a "real-time". Chunked publishing, however, may skew election's outcome as the voters are likely to take into account a momentary standing and to support either a winner or a loser in the moment, which is similar to the ranking influence[6]. On the other hand, some electoral designs may incorporate such influence on purpose.

Proof of Work technique is notorious for using excessive amounts of energy in cryptocurrencies' mining[7], and yet cryptocurrency does not have any intrinsic value to justify such energy feast[8], standardized interest-free electoral process could perfectly do.

Not only could Proof of Work mechanism allow zero-trust elections, it would also readily expose a measure of protection from tampering. Every record of the vote is expected to encompass a "nonce" – a binary sequence that could only be obtained by performing a lot of cryptographic computations (work). The nonce is such that when the vote record is fed to a hash function, a digest that function produces indicates the amount of work originally performed. For instance, the more leading zero bits the digest has - the more work/energy was invested. In fact, we know how much work exactly, provided that the hash function is agreed upon. If an adversary wishes to change the vote record, they would have to re-invest that same amount of work. Maybe they could tamper with a few records, but computational capability combined across all the voters should prevail, so the election outcome would be backed by some known amount of work, or an electric energy figure to that effect. On the other hand, tampering on the large scale can be potentially detected by spikes in energy consumption, corresponding to the backing figure in question.

There would be three roles engaged in the proposed voting protocol, and their characteristics are as follows:

1. An authority – electoral commission, governmental portal or similar. We may have no better option than trusting the authority to define voters. The authority can electronically sign a one-time-pass (unique but not personally identifiable token) and share it with the voter as an authorization (mandate) to vote. The authorization identifies the voter in the election and gets published in the end. Anyone could verify the signature on such authorization with a public key published on behalf of the authority.

2. A platform outside the authority's control. We need such platform to reliably sever the link between the authorization to vote (personally identifying the voter to the issuing authority) and the vote itself. By the end of an election two unrelated lists get published: the list of voted authorizations and the list of casted votes. The platform's software must be open-source or publicly available otherwise. Chunked publishing of the lists during the election may be an option as well.

3. A voter casting their vote runs some open-source software on their device. This software provided by the platform spends some time calculating cryptographic digests from the following input: the vote, a digital receipt (see below), and the nonce. Some work testing millions of nonces will be done to find the one yielding a digest of a certain form. The authorization and corresponding vote(s), the receipt, and the nonce will then be sent to the platform for checking and publication.

No party is trusted, and every possibility to cheat must be explored and addressed. For example, authority can use bots to vote with forged authorizations. To deal with such scenario Completely Automated Public Turing test to tell Computers and Humans Apart (CAPTCHA) may be employed. Generally, interactive challenges – CAPTCHA or a specialized game[9] – would also buy some time for the voter's device to perform more work.

The authority could conspire with some voters and have them sending data blocks with spurious votes prepared in advance – before the election time frame. To tackle this scenario the platform issues a unique stamp – part of the digital receipt. Since the receipt is included in the data block protected by the work, performing the work in advance will not be possible.

Symmetrically, to ensure the platform couldn't prepare fraudulent data blocks in advance each voter generates own unique stamp – also part of the digital receipt. Later on, any vote on the published list could be found and verified by the voter using that receipt. Obviously, the voters can do it independently – without any interference from the authority.

A conspiracy between the authority and the platform can create an opportunity to fraudulently exercise legitimate authorizations to throw in votes and then reject legitimate voters; or to de-anonymize the votes. Making the platform's operations easily observable with the open source software and slicing the platform's architecture across multiple sites can help to mitigate this risk. Continuous (chunked) publishing would also promote observability and tackle throw-ins.

Finally, organized sabotage by the voters remains a problem. For instance, thousands of voters may falsely claim denial of service with a view to derail elections. It is unlikely, however, that any considerable slice of electorate would go this rogue. On the bright side, voter's receipt is designed in a way that s/he would not be able to claim a published voting record from another voter, yet be able to prove the ownership of their own record – a property known as collection accountability[10].

The following diagram outlines high-level architecture of the voting process:

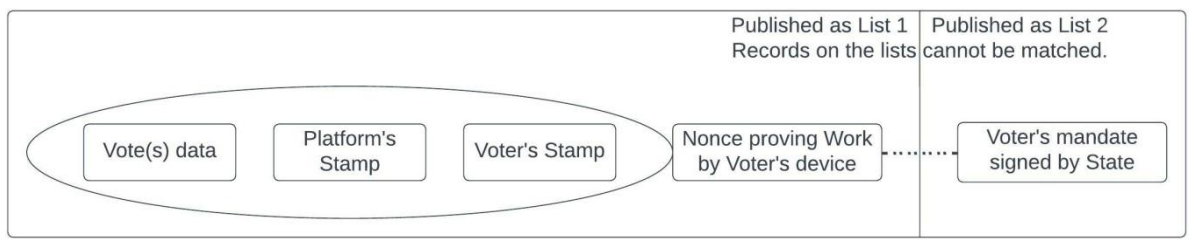

**Figure 1 | A data block en route from a Voter's device to the platform.** Stamps ensure the nonce proves the work within the voting period; proof of that work deters tampering at scale. Voter's stamp makes published votes end-to-end verifiable[10]. Voter's mandate ensures that the voter voted once and was authorized. The platform splitting the data into two published lists ensures secrecy. No personally identifiable data are disclosed.

## Generalization

When blockchain fuelled cryptocurrency had emerged many voices heralded the beginning of a new era where the role of State may wither away. Likewise, it would be interesting to explore philosophical implications of a potential energy-based societal consensus.

The primary role of the state is to enforce "social contract". For that, the state must retain monopoly on money, coercion, and identification of its citizens. The system of political parties, however, may be affected as electronic plebiscite will be available for key decisions that are now made in line with specific views of the political parties. Instead, the parties could get less political so to speak and differentiate by more execution-related metrics.

From economy of scale standpoint a single completely open platform as described shall suffice to serve all the countries. It would only be logical to foster such platform as an international nonprofit organisation, under United Nations umbrella for example. In fact, the very recognition of a country as democratic may become contingent on the usage of the platform as opposed to a conventional (less transparent) ballot.

## Adaptive Proof of Work

On its evolutionary timeline, the Proof of Work algorithm became widely known as a deterrent-style protection from tampering with data blocks transacting cryptocurrency[5]. Fundamentally, it relies on the properties of the well-known cryptographic hash functions (e.g. md5, sha512 or Blake2 to name a few)[11].

A hash function takes a bock of data of an arbitrary length as its input and produces a fixed length digest (also called hash) as its output. The properties of interest are:

1. If input blocks are any different, then respective output digests are most likely different: this is known as collision resistance.

2. If the difference between any input blocks is minimal (i.e. by a single digital bit) then respective digests do not look alike at all: this is known as avalanche effect.

To protect a block of data by "work" we first select a hash function, let it be md5 for the sake of an example. Secondly, we add a string of bits to the block we aim to protect. That string is called nonce because it is always a number used once just to make a unique input to the hash function. We shall now "work" calculating md5 digests cyclically, trying every possible nonce through the cycles, until we come across a digest of some required form. And we are at liberty to stipulate that requirement. One common showcase is to seek a digest where all bits are zeros. Md5 digest is of 128 bits' fixed length. The function properties guarantee that every bit of the digest appears random, so the probability of any specific configuration of 128 bits is $½^{128}$ – this is the same chance as in a tossed coin landing heads 128 times in a row. In other words, in order to get zero digest one would have to perform $2^{128}$ iterations on average. Apparently, that is infeasible amount of work.

The argument above would also be valid to show that if only the first 16 bits of the digest were zero, then $2^{16} = 65536$ iterations must have occurred on average. In fact, we don't even have to fix that number of leading bits. One voter can use a laptop and be prepared to wait a bit, so the digest of their data block could have 24 leading zero bits, in which case we would conclude that they recalculated the hash about 16 million times ($2^{24}$). Another voter can use a mobile phone (weaker CPU compared to a laptop) or just be impatient to wait longer than absolutely necessary, and their block's digest could only have 19 leading zero bits, telling us that they went through half a million of the cycles. Generally, the voting software could adapt amount of the work/energy invested to the individual circumstances of every voter.

In contrast to the blockchain framework, where some feasible amount of work is distributed across numerous units of specialized hardware processing one same data block concurrently, electronic voting would result in a greater number of weaker protected data blocks stored at and published by the platform. It would then be trivial to calculate total amount of work during the election in terms of the number of hashing cycles performed by the millions of voters and to assess proportion of the votes that could have been forged by a certain computational power.

## Summary

Cryptography still has non-realized potentials that are capable to materially change our society. Zero-trust plebiscite may emerge as one of those, offering more secure electoral process and, eventually, streamlining the governance. It would have to be three to "tango": the two are a state and its citizens, and the third is an outsider taking away the agency conflict from the state. Distributed computational work is proposed to deter data tampering (as in the proof of work mechanism). If this proposal is adopted, governance landscape will likely change as readily available referendum deprecates the need for static differentiation across political parties.

## Acknowledgement

Many thanks to Philip Stark at Berkeley for many great references to active research in the field.